\documentclass{icrc2009}

\usepackage{graphicx}   
\usepackage[caption=false]{caption}    
\usepackage[font=footnotesize]{subfig} 
\usepackage{fixltx2e}
\usepackage{url}

\newcommand{\shorttitle}[1]%
{\markboth{Proceedings of the 31\MakeLowercase{$^{st}$} ICRC, {\L}\'{o}d\'{z} 2009}{#1} }
\newcommand{\etal}{\MakeLowercase{\textit{et al.}}} 

\begin{document}

\title{Observation of selected SNRs with the MAGIC Cherenkov Telescope}

\author{\IEEEauthorblockN{E. Carmona\IEEEauthorrefmark{1},
			  M. T. Costado\IEEEauthorrefmark{2}\IEEEauthorrefmark{4},
                          L. Font\IEEEauthorrefmark{3} and
                           J. Zapatero\IEEEauthorrefmark{3} 
			   on behalf of the MAGIC Collaboration}
                            \\
\IEEEauthorblockA{\IEEEauthorrefmark{1}Max-Planck-Institut f\"ur Physik, Munich, Germany}
\IEEEauthorblockA{\IEEEauthorrefmark{2}Instituto de Astrof\'{\i}sica de Canarias, E-38205 La Laguna, Tenerife, Spain}
\IEEEauthorblockA{\IEEEauthorrefmark{4}Departamento de Astrof\'isica, Universidad, E-38206 La Laguna, Tenerife, Spain}
\IEEEauthorblockA{\IEEEauthorrefmark{3}Universidad Aut\'onoma de Barcelona, Spain}}

\shorttitle{Observation of various SNRs}

\maketitle

\begin{abstract}
Supernova Remnants (SNRs) are believed to be the acceleration sites of
galactic cosmic rays. As such they are expected to produce Very High
Energy (VHE) gamma-rays through hadronic and/or electromagnetic
scenarios, hence they are natural targets for observations with
ground-based Imaging Atmospheric Cherenkov Telescopes
(IACTs). Currently, VHE emission has been detected from several SNRs,
making them one of the most abundant types of established galactic VHE
sources. The MAGIC telescope, located in the Canary island of La
Palma, has been performing observations of several SNRs over the last
two years. Parameters like age, distance, radio flux, or possible EGRET
association (i.e., criteria matching those already used for previous
successful IACT SNR detections), were used to select candidate
targets. Here we summarize the results of the past two years of
observations.
\end{abstract}



\section{Introduction}

It has been believed for long time that supernova (SN) explosions and
their remnants (SNRs) are the places where the acceleration of
galactic cosmic rays takes place. Energetic arguments as well as the
diffuse shock acceleration mechanism applied to young SNRs support
this idea. A SN explosion every 30 years is enough to balance the
escape losses of cosmic rays in the Galaxy.


During the past years, IACTs have confirmed that SNRs are sources of
VHE gamma-rays. Two scenarios can explain the production of the
gamma-rays. In the electromagnetic scenario, accelerated electrons can
up-scatter low energy photons through the inverse Compton
mechanism. In the hadronic scenario, accelerated nuclei can interact
with matter or radiation producing neutral mesons that will decay into
gamma-rays.

\subsection{The MAGIC telescope}

\textbf{MAGIC} (\textit{Major Atmospheric Gamma-ray Imaging
  Cherenkov}) is a ground based gamma-ray telescope located on the
Canary island of La Palma\footnote{The MAGIC telescope is operated on
  the island of La Palma by the MAGIC Collaboration in the Spanish
  Observatorio del Roque de los Muchachos of the Instituto de
  Astrof\'isica de Canarias}. It is a new generation IACT with a
 trigger threshold of 55~GeV, and an energy detection range from
 60~GeV to 10~TeV, overlapping with the upper energy threshold of
 satellites such as Fermi.  The design of the MAGIC telescope was a
 technological challenge
which took the existing technology to its limit~\cite{diseno,
  diseno2}.  One of its main characteristics is its octagonal
parabolic reflector of 17~m of diameter resulting in an area of
240~m$^{2}$, capable of obtaining three times more light than a
conventional IACT of 10~m. For energies above 150~GeV, the telescope
angular and energy resolutions are $\sim$0.1$^\circ$ and $\sim$25\%\,
respectively \cite{albert2008}. Besides this, in April 2007 its data
acquisition system was upgraded with multiplexed 2~GHz FADCs which
improved the timing resolution of the recorded shower
images. Accordingly, the integral sensitivity of MAGIC improved
significantly from 2.2\% to 1.6\% of the Crab Nebula flux above
270~GeV for 50 hours of observation \cite{aliu2009}.

The MAGIC collaboration has finished the
construction of a second telescope, MAGIC-II, similar to MAGIC with
improved technology. Located at a distance of eighty five metres from
the first telescope, it will be fully operational in the second half
of 2009. The operation of both telescopes in stereoscopic mode will
allow us to improve the spatial resolution and sensitivity of the
MAGIC experiment.

\section{SNRs Observed by MAGIC}

During 2007 and 2008 the MAGIC telescope has performed observations of
different SNRs. The data include a deep observation of the Tycho SNR
and shorter observations of various radio selected SNRs.

\subsection{Tycho SNR (G120.1+1.4)}

The Tycho SNR is one of the best known and most studied SNRs. It is a
shell-type SNR (well defined in radio and X-Ray) which was formed
from, most likely, a Ia supernova explosion in 1572. This young SNR is
a bright X-ray source with a similar diameter of 8' in both X-rays and
radio. The spatial structure of Tycho has been studied by Chandra and
XMM-Newton, providing constraints to the allowed explosion
models. Although the age of the SNR is known with precision, the
distance is not well known. Distance estimates vary between
2.2~kpc~\cite{Albinson} and 4.4~kpc~\cite{Schwarz}. Most accepted
values are in the range 2.3-2.8~kpc. However,  recent measurements
based on the light echo~\cite{echo} suggest that a
larger distance of 3.8~kpc could also be possible.

VHE gamma-ray emission from Tycho is predicted by the non linear
kinetic theory of Cosmic Ray acceleration in SNRs~\cite{Volk}~\cite{Volk2}. In such
models, the dominating mechanism for gamma-ray emission is $\pi^0$
decay rather than inverse Compton. However, spatial correlations in
various SNRs between hard X-rays and VHE gamma-rays favour the idea
that energetic electrons are responsible for the gamma-ray
emission~\cite{RX-J1713.7}, \cite{RX-J0852.0}. The detection of VHE
gamma-rays from the Tycho SNR above several TeVs would help to clarify
the nature of the mechanism responsible for the VHE gamma-ray
emission.

Tycho has been observed at VHE gamma-rays by the
Whipple~\cite{whipple} telescope and the HEGRA telescope
array~\cite{hegra}, however none of them found VHE gamma-ray
emission. In the case of HEGRA, a search for TeV
gamma-radiation from Tycho SNR was performed over 2 years
(1997/98). No evidence for such emission was found and a $3\sigma$
level upper limit was estimated: 3.3\% Crab at 1 TeV ($5.79 \times
10^{-13}\,$ph$\,$cm$^{-2}\,$s$^{-1}$).

\subsection{Radio Selected SNRs}

The Green Catalog of SNRs \cite{green} contains 265 Galactic SNRs plus
some possible candidates. It includes information about the flux and
spectral index at 1~GHz, summarizing the information scattered over
hundreds of publications and catalogs. This information has been used
to select a list of SNRs whose parameters are similar to those of 6
well identified SNRs that are also known to be VHE gamma-ray emitters:
IC443, RXJ0852-4622, RCW86, RXJ1713-3946, W28 and Cas A. From the known
SNRs with VHE emission, we defined the following selection
criteria\footnote{Some of the parameters are unknown or have large
  uncertainties. The selection criteria are not a strict rule for
  selection but a guideline.} to find targets for our observations:

\begin{itemize}
\item {\bf Flux at 1 GHz $\geq$ 49~Jy}
\item {\bf Radio Spectral Index ($S_\nu \propto \nu^{-\alpha}$) $\leq$ 0.6}
\item {\bf Distance $\le$ 7~kpc}
\item {\bf Age $\le$ 50000~yr}
\end{itemize}

From all the sources contained in the Green Catalog, a total of 37
candidates met the selection criteria.  Only 25 of these sources are
observable by MAGIC with a zenith angle below 50$^\circ$. From these 25
sources some where removed because they have already been detected in
gamma-rays (Crab, IC443, ...), were previously observed (Tycho) or
were covered by the H.E.S.S. galactic scans. Finally we
selected the 9 sources shown in table~\ref{selection_criteria} as
interesting targets to perform observations with MAGIC.

\begin{table}[!h]
\caption{Parameters of the selected SNRs.}
\label{selection_criteria}
\centering
\begin{tabular}{c|c|c|c|c|c}    
\hline
Source & Diameter & Distance & Age & Flux & Radio \\
 & (arcmin) & (kpc) & ($10^3$~yr)& 1~GHz & Spec.\\
 & & & & (Jy) & Index\\
\hline
HB-9 & 130 & 1 & 7.7 & 110 & 0.6\\
W51 & 30 & 6 & 30 & 160 & 0.3\\
CTB-80 & 80 & 2 & 100$^2$ & 120 & --\\
W63 & 80 & 1.6 & -- & 120 & 0.5\\
W66 & 60 & 1.5 & 50 & 340 & 0.5\\
HB-21 & 105 & 0.8 & 19 & 220 & 0.4\\
G85.4+0.7 & 24 & 3.8 & 6.3 & -- & 0.5\\
G85.9-0.6 & 24 & 5 & 4 & -- & 0.5\\
CTB-104 A & 80 & 1.5 & -- & 65 & 0.4\\
\hline
\end{tabular}
\end{table}

\section{Observations and Results}

All observations were performed in false-source tracking mode, named
Wobble mode \cite{wobble}, with two directions at a distance of 24'
and opposite sides of the source position. This technique allows for a
reliable estimation of the background with no need of extra
observation time.

Data analysis was carried out using the standard MAGIC analysis and
reconstruction software chain, which proceeds in several steps:
calibration \cite{calibration}, image cleaning procedure
\cite{aliu2009} and parametrization with Hillas parameters
\cite{hillas}. The signal-to-noise maximization is achieved using a
multidimensional classification procedure based on the Random Forest
method \cite{random}, where a hadron likeness measure
(\emph{hadronness}) is computed for each event based on the image and
time parameters.  A skymap has been obtained for events surviving a
\emph{hadronness} cut, to search for signal in the field of view
around the source. We have also obtained upper limits to the VHE flux
for all sources when no positive detection was found. These upper
limits have been computed using the Rolke method \cite{rolke} at a
95\% confidence level and they take into account a 30\% of systematic
uncertainties in the flux level.

\subsection{Tycho SNR}

MAGIC observations of Tycho were performed between July and November
2007. The zenith angle of the observations ranged between 35$^\circ$
and 50$^\circ$. Part of the data were taken under moderate moonlight
conditions (40\%). The total observation time is 69.4 hours after
quality cuts (mainly rejection of bad weather runs).  During moonlight
observations the trigger discriminator threshold varied between 15 and
25 arbitrary units to keep a low rate of accidental triggers. The
effect of the higher discriminator threshold is negligible compared to
the rise due to the medium to high zenith angle. Dark and Moon data
are analyzed together using the standard MAGIC analysis. The recorded
images were cleaned using time image cleaning with boundary cuts of 8
and 4 photoelectrons (typical values for Dark observations are 6 and
3). The 8' diameter of the Tycho SNR is very close to the angular
resolution of MAGIC (6'), so the source is very close to point-like
for MAGIC.

\begin{figure}[!t]
\centering
\includegraphics[height=0.23\textheight]{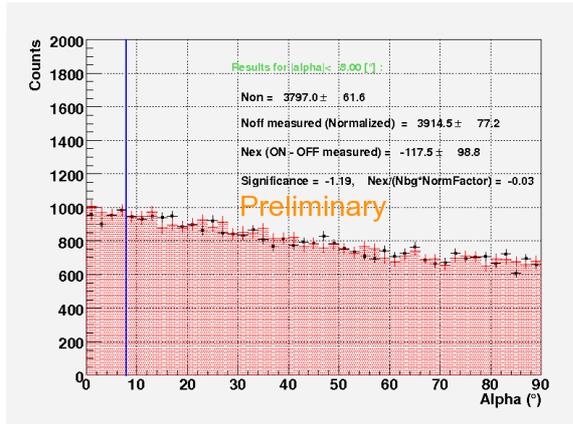}
\caption{Alpha distribution }
\label{fig: tychoAlpha}       
\end{figure}

The so-called $\alpha$ distribution for the source and the anti-source
positions are shown in Figure~\ref{fig: tychoAlpha} for an energy
threshold of 350 GeV (Size cut of 300 and reconstructed energy cut
300~GeV).

No positive evidence for VHE gamma-ray emission from Tycho was
found. For an energy threshold of 350~GeV we obtain an integral upper
limit (3$\sigma$) for point-like source emission (valid hypothesis for
Tycho, given its size) of
$1.86\times10^{-12}\,$ph$\,$cm$^{-2}\,$s$^{-1}$, which corresponds to
$\sim$2\% of the Crab spectrum at the same energy. For an energy
threshold of 1~TeV the corresponding integral upper limit (3$\sigma$)
is $2.95\times10^{-13}\,$ph$\,$cm$^{-2}\,$s$^{-1}$ ($\sim$1.7\% Crab),
50\% better than that obtained by the HEGRA telescope array. With a
confidence level of 2$\sigma$ the corresponding upper limits at
350~GeV and 1~TeV are respectively
$7.04\times10^{-13}\,$ph$\,$cm$^{-2}\,$s$^{-1}$ ($\sim$0.7\% Crab) and
$1.16\times10^{-13}\,$ph$\,$cm$^{-2}\,$s$^{-1}$ ($\sim$0.7\% Crab).

\subsection{Radio selected SNRs}


All selected sources were observed for a time varying between $\sim$5 and
$\sim$10~hours, depending on the SNR. Although this time is clearly
small (specially for extended sources and off-axis) it could trigger
deeper observations in case that a signal hint was found. Table
\ref{table_snr} shows some details of the observations: central
position of the observations (not always the center of the SNR), date
of the observations and effective time after quality cuts (bad weather
rejection). The observations were done under Moon conditions. Only
W66 ($\gamma$-Cygni) was observed under Dark conditions. The images
were cleaned using time image cleaning with boundary cuts of 8 and 4
photoelectrons for the Moon observations and 6 and 3 photoelectrons
for the Dark observations of W66.

Assuming a conservative sensitivity of 2\% Crab (5$\sigma$, 50 hours)
for the Crab Nebula spectrum, we can expect to find a point like
source at $5\sigma$ level in $\sim$8 hours for a 5\% Crab flux or in
$\sim$2 hours for a 10\% Crab flux, if the source was point-like and
located at the center of the two wobbling positions used in each
observation. The small trigger area of MAGIC ($\sim 0.9^{\circ}$
diameter) makes things more difficult when the source is not located
at the center of the observations or when the source is extended. For
a point-like source located $0.5^{\circ}$ from the center of the
observations the times needed for a 5\% Crab flux and a 10\% Crab
detection at $5\sigma$ level would be $\sim$30~hours and $\sim$8~hours
respectively.

\begin{table}[!h]
\caption{Some parameters of the observations performed on the 9 radio
  selected SNRs. All observations are done in wobble mode in Moon
  time, except W66.}
\label{table_snr}
\centering
\begin{tabular}{c|c|c|c|c}    
\hline
Source & RA  & DEC & Date & Eff. Time \\
 & (h) & ($^o$) & (Month/Year) & (h) \\
\hline
HB-9 & 5.02 & +46.67 & 08-09/08 & 6.35 \\
W51 & 19.40 & +14.51 & 07-08/08 & 7.38 \\
CTB-80 & 19.89 & +32.88 & 07-09/08 & 5.06 \\
W63 & 20.32 & +45.7 & 11-12/08 &  5.36 \\
W66 & 20.36 & +40.26 & 06-09-11/08 & 11.4 \\
HB-21 & 20.75 & +50.6 & 08-09-11/08 &  7.90 \\
G085.4+0.7 & 20.85 & +45.37 & 06-07/08 & 5.21 \\
G085.9-0.6 & 20.98 & +44.9 & 06-07/08 & 6.28 \\
CTB-104 A & 21.49 & +50.8 & 07-08/08 & 7.97 \\
\hline
\end{tabular}
\end{table}

After analysis of all sources, we found no significant ($>5\sigma$) gamma-ray
emission from any of them. We have obtained integral upper limits for
the VHE flux coming from a point-like source located at the center of
the wobble observations. The corresponding
integral upper limits ($3\sigma$) above an energy threshold of 270~GeV
are shown in table~\ref{upper_limits}.

\begin{table}[!h]
\caption{Upper Limits for a point-like source located at the center of
  the observations (not necessarily the center of the SNR) for
  energies above 270~GeV.}
\label{upper_limits}
\centering
\begin{tabular}{c|c|c}    
\hline
Source & Integral Flux UL & Flux U.L.\\
& (ph~cm$^{-2}$~s$^{-1}$) & Crab units\\
\hline
HB-9 & $1.60\times10^{-11}$ & 11\%\\
W51 & $1.26\times10^{-11}$ & 9\%\\ 
CTB-80 & $3.56\times10^{-11}$ & 25\%\\
W63 & $3.36\times10^{-11}$ & 24\%\\
W66 & $6.68\times10^{-12}$ & 5\%\\
HB-21 & $7.88\times10^{-12}$ & 6\%\\
G085.4+0.7 & $2.58\times10^{-11}$ & 18\%\\
G085.9-0.6 & $2.10\times10^{-11}$ & 15\%\\
CTB-104 A & $1.40\times10^{-11}$ & 10\%\\
\hline
\end{tabular}
\end{table}

The skymaps obtained for every source are shown in figure~\ref{fig:
  MultiSkymap} where no significant point-like gamma-ray excess has
been found. For extended sources, the significance is reduced in our
observations because of the small trigger area of MAGIC which makes
difficult the detection of sources of various tenths of degree. In the
case of W51, the recently reported VHE emission from J1923.0+1411 by
MILAGRO~\cite{milagro} is coincident with an area of higher
significance in the rim of the shell (see Figure~\ref{fig:
  MultiSkymap}). H.E.S.S. reported~\cite{hess-w51} a 3\% Crab flux
from this source which is well below our upper limit (9\% Crab).  The
small observation time (7.4~hours), the size of the source and its low
flux made it not possible to have a significant signal in our
observations and only a hint of the source can be seen as the
$\sim3\sigma$ significance area in the skymap.

\begin{figure*}[th]
  \centering
  \includegraphics[width=6in]{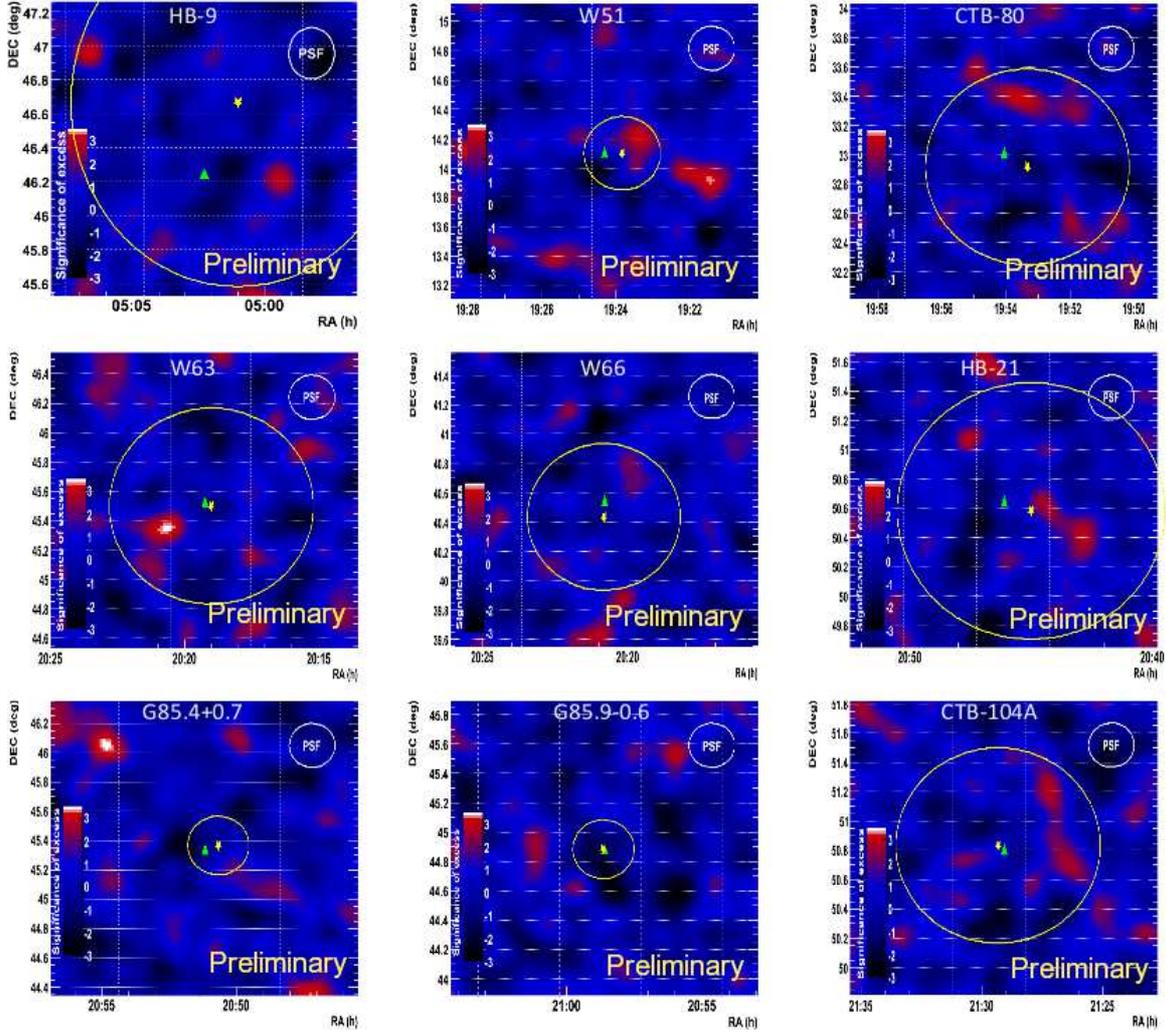}
  \caption{Skymaps of all 9 radio selected SNRs. The yellow star marks
    the center of the SNR and the circle is the approximate boundary
    of the radio shell (data from the Green Catalog). The green triangle
    marks the center of the wobble observations. Starting from
    top-left and to the right: HB-9, W51, CTB-80, W63, W66, HB-21,
    G085.4+0.7, G085.9-0.6 and CTB-104 A.}
  \label{fig: MultiSkymap}
\end{figure*}

\section{Conclusions}
No significant VHE gamma-ray emission has been found for Tycho and the
radio selected SNRs. In the case of Tycho, VHE gamma-ray emission is
predicted at a level that should have been detected in our
observations if the distance is below $\sim3.5$~kpc. The distance to
the SNR is still uncertain although most measurements lie below
3.6~kpc. Our result may indicate that the distance could be larger
than it is usually believed. In the case of the radio selected SNRs,
the computed upper limits range from 5\% to 25\% Crab, assuming a
point-like source at the center of the observations. In addition, the
skymaps show no significant ($>5\sigma$) evidence of VHE emission
coming from  point-like or small size sources located inside the
shell of the SNRs.




\end{document}